\documentclass[preprint,12pt]{elsarticle}

\usepackage{amssymb}
\usepackage{amsmath}
\usepackage{lineno}
\usepackage{listings}

\usepackage[utf8]{inputenc}
\usepackage[T1]{fontenc} 
\usepackage{commath}

\usepackage{setspace}
\usepackage{graphicx}
\usepackage{siunitx}
\usepackage{gensymb}
\usepackage{color}

\journal{Ultramicroscopy}

\begin{document}

\begin{frontmatter}

\title{Free, flexible and fast: orientation mapping using the multi-core and GPU-accelerated template matching capabilities in the python-based open source 4D-STEM analysis toolbox Pyxem}


\author[p:mpie]{Niels Cautaerts}
\author[p:cam]{Phillip Crout}
\author[p:trond]{H{\aa}kon Wiik {\AA}nes}
\author[p:man]{Eric Prestat}
\author[p:mpie]{Jiwon Jeong}
\author[p:mpie]{Gerhard Dehm}
\author[p:mpie]{Christian H. Liebscher}

\address[p:mpie]{Max-Planck-Institut für Eisenforschung GmbH, Max-Planck-Straße 1, 40237, Düsseldorf, Germany}
\address[p:cam]{Department of Materials Science \& Metallurgy, University of Cambridge, 27 Charles Babbage Road, Cambridge, United Kingdom}
\address[p:trond]{Department of Materials Science and Engineering, NTNU, Alfred Getz vei 2, N-7491,Trondheim, Norway}
\address[p:man]{Department of Materials, University of Manchester, The Mill E2, M13 9PL, Manchester, United Kingdom}


\begin{abstract}
This work presents the new template matching capabilities implemented in Pyxem, an open source Python library for analyzing four-dimensional scanning transmission electron microscopy (4D-STEM) data.
Template matching is a brute force approach for deriving local crystal orientations. 
It works by comparing a library of simulated diffraction patterns to experimental patterns collected with nano-beam and precession electron diffraction (NBED and PED).
This is a computationally demanding task, therefore the implementation combines efficiency and scalability by utilizing multiple CPU cores or a graphical processing unit (GPU).
The code is built on top of the scientific python ecosystem, and is designed to support custom and reproducible workflows that combine the image processing, template library generation, indexation and visualisation all in one environment.
The tools are agnostic to file size and format, which is significant in light of the increased adoption of pixelated detectors from different manufacturers.
This paper details the implementation, validation, and benchmarking results of the method.
The method is illustrated by calculating orientation maps of nanocrystalline materials and precipitates embedded in a crystalline matrix.
The combination of speed and flexibility opens the door for automated parameter studies and real-time on-line orientation mapping inside the TEM.
\end{abstract}


\end{frontmatter}




\section{Introduction}
Scanning nanobeam electron diffraction (NBED) and precession electron diffraction (PED) have been applied for over two decades inside transmission electron microscopes (TEM) for orientation mapping in nano-crystalline materials~\cite{Rauch2008, Rauch2010}.
These methods rely on scanning an electron probe with a small convergence angle on the order of 1~mrad across an electron transparent sample while capturing a nanobeam diffraction pattern image at each scan point.
A common approach to extracting orientations from the diffraction patterns is to use template matching, whereby the pattern is compared to a large library of pre-computed templates of simulated diffraction patterns~\cite{Rauch2008, Rauch2010}.
For sufficiently thin samples and a well aligned electron probe a spatial resolution of around 1~nm can be achieved~\cite{jeong2021automated}.
However, the angular resolution of the technique tends to be limited to about 1\degree~\cite{Zaefferer2011}. 
These limitations notwithstanding, the technique enjoys widespread use as a fast and convenient method for nano-scale orientation mapping and has been extensively utilized to determine grain orientations in nanocrystalline materials~\cite{viladot2013orientation, mompiou2015quantitative, brons2014orientation} even combined with in-situ straining~\cite{kobler2013combination}.

Recently interest in the field of 4-dimensional scanning transmission electron microscopy (4D-STEM) has been reignited thanks to the emergence of fast and direct electron detectors~\cite{ophus2019four}.
With direct electron detectors, high quality diffraction patterns can be collected in a fraction of the time compared to the camera systems that are currently most used.
Additionally, as some of the authors on this manuscript have shown, NBED orientation mapping can also benefit from the improved signal-to-noise characteristics of pixelated detectors~\cite{cautaerts2021investigation, jeong2021automated}.
Improved detector technology results in datasets that are larger and more complex, requiring customizable analysis frameworks.
As the complexity of the analysis pipelines increases it becomes increasingly important for them to be traceable and reproducible.

For these reasons, a fast, scaleable, and flexible template matching workflow was implemented in the open source Pyxem library. 
The implementation makes use of the just-in-time (JIT) Numba compiler~\cite{lam2015numba} for compiling performance critical parts of the code, Dask~\cite{rocklin2015dask} for parallelizing the workload, and CuPy~\cite{nishino2017cupy} for performing the calculations on the GPU.
Pyxem is a free and open source Python library that offers a large number of 4D-STEM data analysis routines~\cite{duncan-n-johnstone-2020-3976823}.
It builds on the open source HyperSpy library~\cite{de2017electron} for microscopy data analysis.
HyperSpy includes a large number of file readers, making workflows in Pyxem file format agnostic.
With Pyxem and HyperSpy, analysis pipelines are built up with a minimal amount of python code in Jupyter notebooks~\cite{kluyver2016jupyter} or scripts; these are transparant and can be made completely reproducible.
As the entire code base is written with Python syntax, development times are rapid and the code can be easily contributed to by non-professional programmers like scientists in the field.

This paper discusses the implementation details of the new template matching algorithm, demonstrates its performance, compares the results to those obtained with a commercial solution, and illustrates with case studies. 

\section{The algorithm and implementation}
\subsection{Simulation of diffraction pattern templates using diffsims}
Indexation of orientations via template matching relies on a library of pre-computed diffraction pattern templates.
The library is calculated with diffsims, a Pyxem sub-library~\cite{diffsims2021}.
Each template is stored as a list of Bragg reflections with reciprocal space coordinates $(k_x, k_y)_i$ and associated intensities $I_i$.
The template library consists of a list of crystal orientations that each have an associated template.
Sampling all of orientation space ($SO(3)$) at small increments requires a very large template library, but the size can be drastically reduced by observing that many orientations produce templates related by a rotation in the imaging plane.
If orientations are represented by Euler angles $(\phi_1, \Phi, \phi_2)$ according to the Bunge convention~\cite{bunge1969mathematische}, then all templates that share the same $\Phi$ and $\phi_2$ are related by in-plane rotation.
Therefore, the list of orientations making up the library must only sample ($\Phi$, $\phi_2$) or equivalently the surface of the unit sphere ($S^2$); the in plane angle $\phi_1$ is set to zero for all the templates and is found during the indexation process.
The list of orientations is further reduced by crystal symmetry to a subsection of $S^2$ representing unique beam directions; all Laue groups are supported in the software.

\begin{figure}[h]
    \centering
    \includegraphics[width=\linewidth]{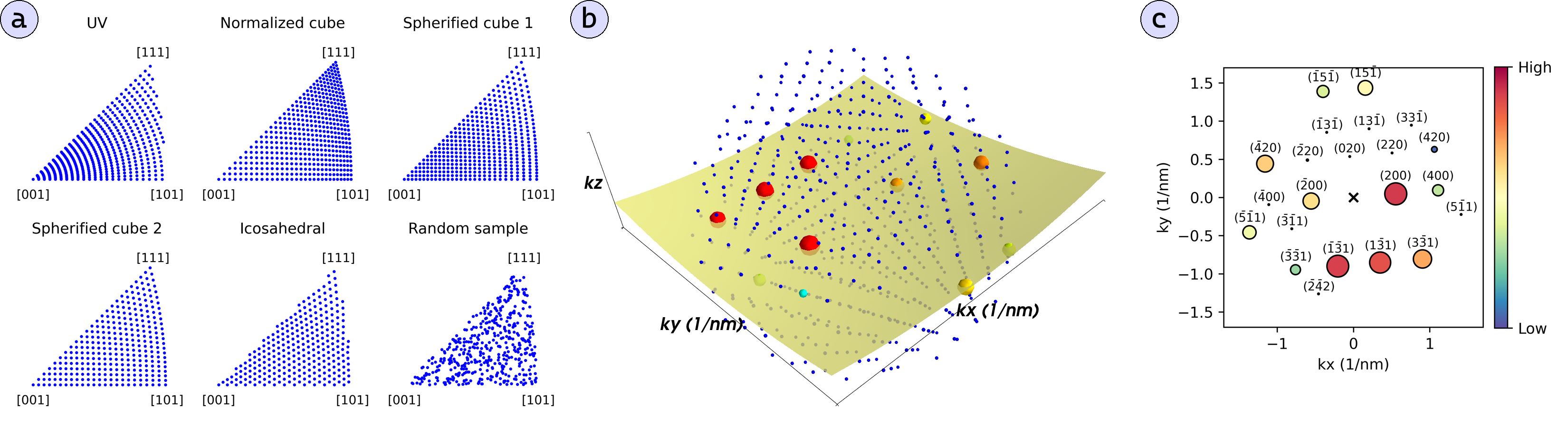}
    \caption{Illustration of the template library procedure in diffsims. a) Various orientation grids for the cubic crystal structure generated with different meshing algorithms. The orientation is represented by the beam directions plot in the stereographic projection. b) The reciprocal space grid of FCC Fe rotated to $\Phi=15$~\degree and $\phi_2=5$~\degree. The Ewald sphere is represented by the yellow surface. Spots closer to the surface are more intense, represented with being larger and more intense colors. c) The projected diffraction pattern with indexed spots. Spots below a minimum intensity threshold were filtered out.}%
    \label{fig:simulation}
\end{figure}

The first step in creating a template library is to form a list of candidate orientations represented by points on the unit sphere surface.
Different sampling schemes were implemented in diffsims as shown in Figure~\ref{fig:simulation}(a); for the rest of this paper the \emph{Spherified cube 2} sampling is used.
When the sampling density is sufficiently high, i.e. when the sampling interval is smaller than the precision of the method, the type of sampling does not have a big influence on the results.

In the second step, a kinematical diffraction pattern is simulated for each orientation on the grid.
A reciprocal space grid within a limiting radius is constructed, rotated by $(0, \Phi, \phi_2)$, and the distance of each grid point to the surface of the Ewald sphere is calculated.
This geometric excitation error is used in a shape factor function, representing the relrods centered on the diffraction spots, to determine the spot intensity.
Multiple shape factor functions are implemented and the user can provide custom functions if desired.
The reflection intensities are determined by the structure factor $F$ through the well known relation $I=FF^*$; finally $I$ is multiplied by the shape factor to obtain the spot intensity.
Structure factors are calculated using a user supplied crystal structure and atomic scattering factors calculated from parametrizations described in literature~\cite{kirkland1998advanced, lobato2014accurate}.
An alternative approach is to set atomic scattering factors to 1, which may be preferred for template matching~\cite{rauch2005rapid} since it increases the weight of faint distant reflections in calculating the best matching template.
The final step to obtaining the 2D diffraction pattern consists of projecting the spots close to the Ewald sphere onto the $k_x-k_y$ plane and discarding the rest.

This process is illustrated in Figure~\ref{fig:simulation}(b), where the large yellow surface represents part of the Ewald sphere and the points represent the reciprocal space grid of face centered cubic (FCC) Fe (serving as a simple prototype for austenitic steel).
Points far away from the Ewald sphere are not excited and are represented with small blue spheres, whereas reflections that are close by and intense are represented as large spheres going from green, to yellow, to red.
The resulting projected pattern is shown in Figure~\ref{fig:simulation}(c); the size and color of the reflections serve to represent their intensity.

The method in diffsims can also be used to simulate beam precession, by adopting the formalisms described in ~\cite{palatinus2019specifics}.

\subsection{Indexation workflow}
Each image in the 4D-STEM dataset must be compared to each template in the library.
Simultaneously, the in-plane angle $\phi_1$ that produces the best match between image and template must be found.
Following the method described in reference~\cite{Rauch2019}, the measure of best fit is taken to be the correlation index as described by equation~\ref{eq:corindex}:

\begin{equation}
\label{eq:corindex}
    Q = \frac{\sum_i P(x_i, y_i) T(x_i, y_i)}{\sqrt{\sum_i P(x_i, y_i)^2} \sqrt{\sum_i T(x_i, y_i)^2}},
\end{equation}

\noindent where $P(x, y)$ represent coordinates in the image and $T(x, y)$ coordinates in the template for all pixels $i$.
In practice, the template is very sparse and only non-zero at coordinates corresponding to diffraction spots, so the sum in the numerator is reduced to a sum over all the template diffraction spots.
The denominator serves the purpose of optionally normalizing $P$ and $T$; these values can be pre-computed and applied to images and templates before indexation, thereby reducing the equation for $Q$ to a dot product.

During the preparation of this manuscript alternative quality metrics such as Pearson's correlation coefficient were considered.
This was motivated by perceived limitations of $Q$, including the fact that weak high-index reflections, which may be more sensitive to small changes in orientation than strong low-index reflections, don't substantially contribute to the dot product.
In addition, erroneous reflections in the template that do not match with a reflection in the image are not penalized, because in calculating $Q$ their intensity is multiplied by the background intensity that is close to zero.
However, alternative quality metrics did not noticeably improve the reliability of the indexation result compared to when $Q$ was used, but they were significantly more computationally expensive.
In practice, carefully tailored image processing combined with $Q$ produced the best results.
Processing steps included applying a gamma correction, which raised all pixel values to a power $<1$, to the experimental diffraction patterns overcomes the first limitation by increasing the relative intensity of weak reflections.
Subtracting a constant from all the pixels in the image helps to overcome the second limitation of $Q$, as erroneous positive reflections in the template are multiplied by a negative value from the background, thereby penalizing $Q$.

\begin{figure}[h]
    \centering
    \includegraphics[width=\linewidth]{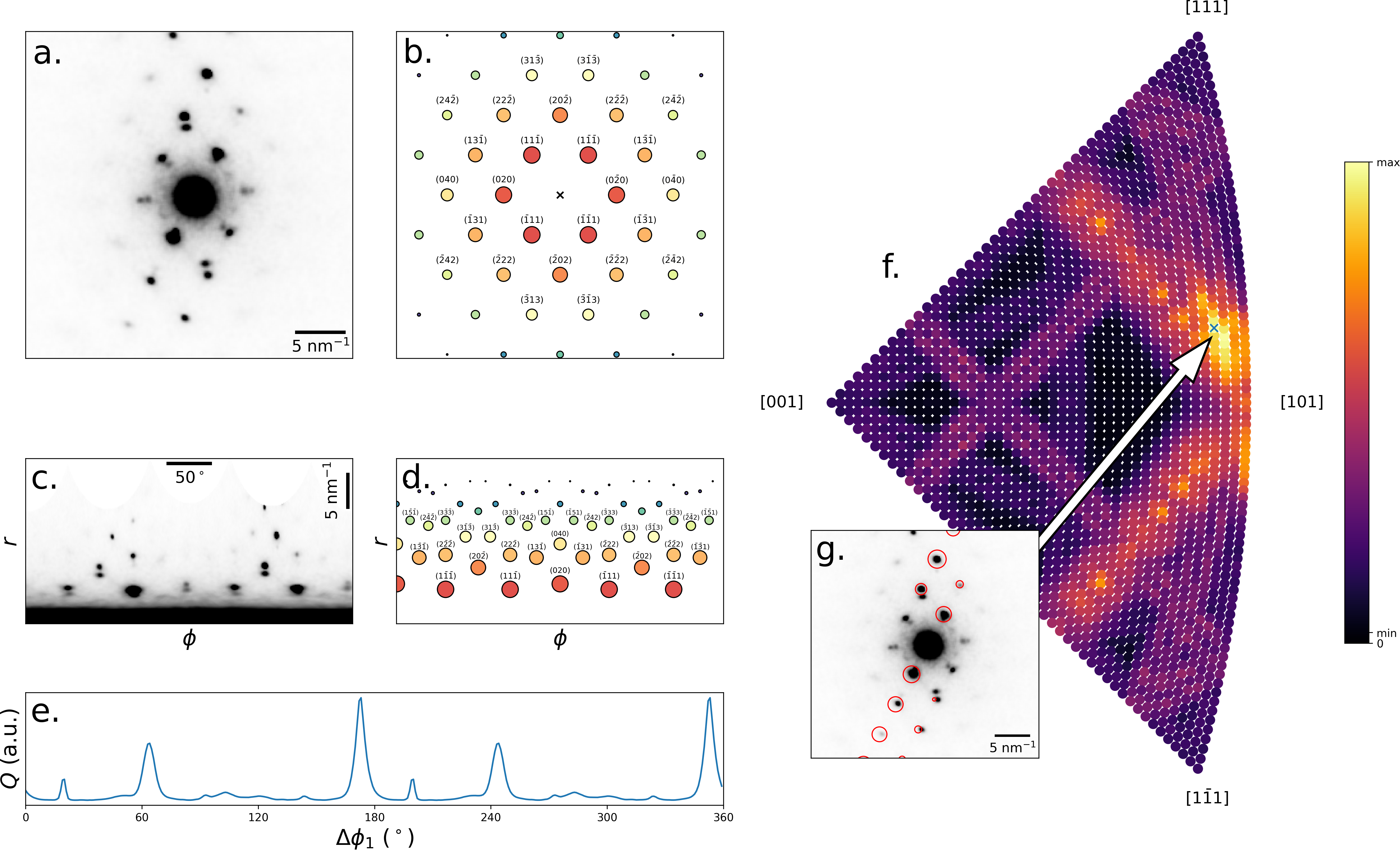}
    \caption{Illustration of the indexation procedure. (a) A diffraction pattern and (b) a template are converted to polar coordinates as shown in (c) and (d) respectively. (e) The template is shifted across the polar image and the correlation index $Q$ is calculated at each position. The process is repeated for all templates and the mirror images of each template and the maximum correlation index and the angle at which it occurs is recorded for each. (f) The best correlation index for each template is mapped to each orientation and plot on the stereographic projection. Better matching templates are brighter, and the template with the highest correlation is taken to be the solution. The minimum and maximum correlation values are indicated on the color bar; precise values of $Q$ are arbitrary and depend on image preprocessing and simulation parameters. (g) The reflections in the best matching template is represented with red circles on top of the experimental pattern.}%
    \label{fig:indexationprocess}
\end{figure}

Different approaches can be used to optimize the in-plane rotation angle $\phi_1$ when comparing a template to an image.
Currently Pyxem implements a direct approach which is illustrated in Figure~\ref{fig:indexationprocess} (a)-(e).
The image (a) and template (b) are both converted to polar coordinates, (c) and (d) respectively.
This is computationally efficient.
However, it is important that the center of the direct beam is used as the origin for the transformation.
The optimal $\phi_1$ is found by sliding the template across the image, with wraparound, along the azimuthal axis $\phi$ and calculating $Q$ at each shifted position as shown in Figure~\ref{fig:indexationprocess} (e).
The diffraction spot coordinates of the templates $(r_i, \phi_i')$ are rounded to the nearest integer so that they correspond directly to coordinates of pixels $P'(r_i, \phi_i')$ that are used to evaluate equation~\ref{eq:corindex}.
The step size in $\phi$, and thus the resolution of the in-plane angle, is determined by the angular sampling used during the polar transform.
It corresponds to 360\degree divided by the number of pixel columns in the polar image.
In the example of Figure~\ref{fig:indexationprocess} (c), the polar image is 360 pixels wide so $\Delta \phi_1 = 1$\degree.
The maximum correlation index and the angle at which it is achieved are stored, and the process is repeated for all templates in the library.
The best correlation index for each template can be plotted on the stereographic projection as shown in Figure~\ref{fig:indexationprocess} (f).
Note that the region in the stereographic projection is twice as large as the original stereographic projection representing the templates, since mirror images of all templates also represent unique orientations and must be considered.
Mirrored templates are generated by reflecting spots across the $y=0$ plane.
The experimental pattern is then indexed by selecting the template and corresponding in-plane angle with the highest correlation index as shown in Figure~\ref{fig:indexationprocess} (f) and (g).
Alternatively, at the cost of some additional computation, the list of correlations can be sorted and the top $N$ best matching templates can be queried instead of only the best one. 

The time complexity of the direct approach to match one template to one image is $O(s\phi)$ with $s$ the number of reflections in the template and $\phi$ the width of the polar image.
If $s$ becomes very large and the template approaches a dense image, i.e. $s\approx r \phi$ with $r$ the height of the polar image, then the complexity becomes $O(r \phi^2)$.
In this case a fast Fourier transform (FFT) based cross-correlation method with time complexity $O(r\phi\log\phi)$ may be preferred~\cite{tzimiropoulos2010robust}.
This method no longer depends on the number of diffraction spots in a template, but benchmarks showed that a few hundred diffraction spots per template are necessary before the cross-correlation approach outperforms the direct approach.

\begin{figure}[h]
    \centering
    \includegraphics[width=\linewidth]{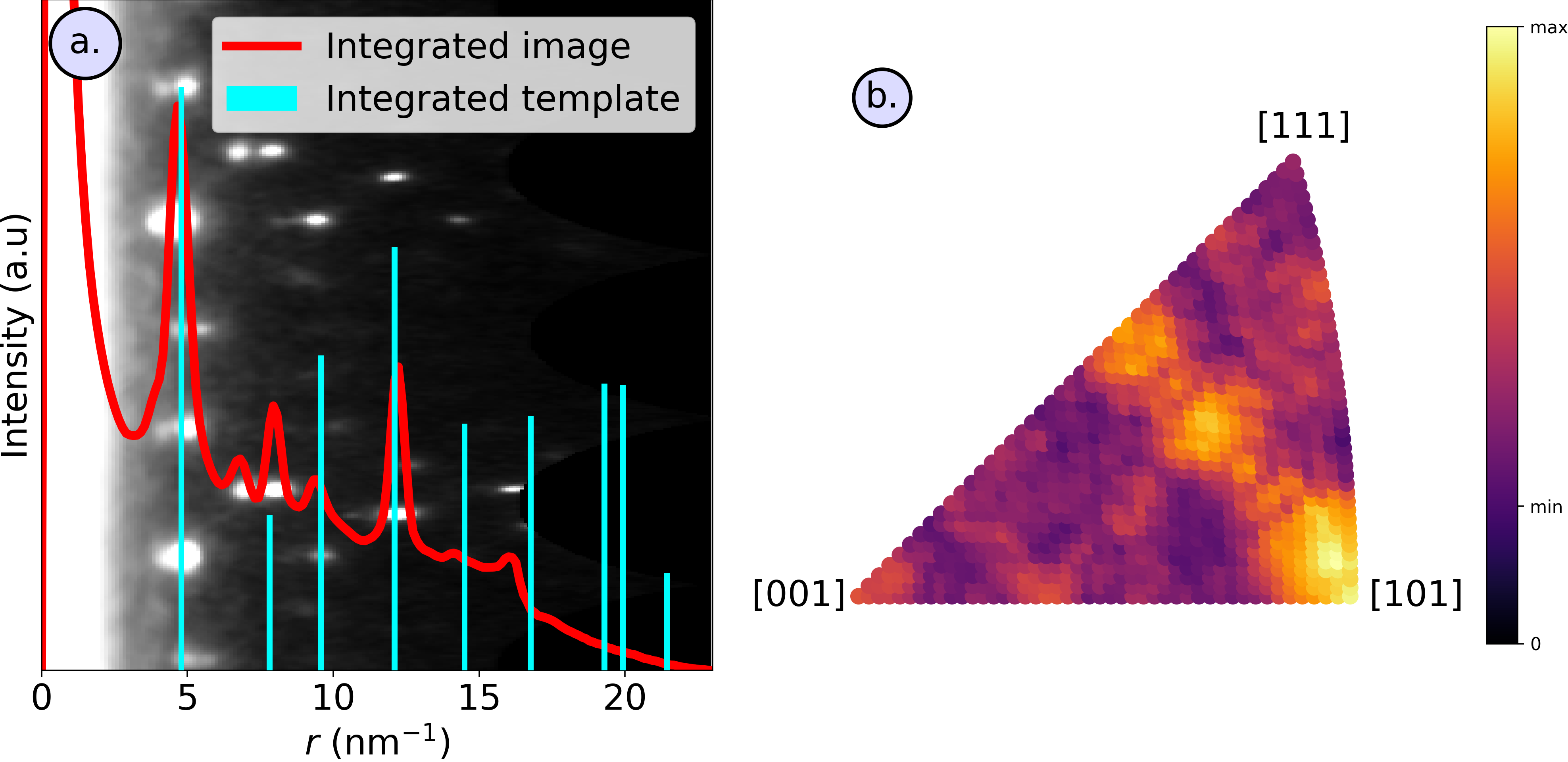}
    \caption{Illustration of the fast matching between integrated images and templates on the same image and template library as in figure~\ref{fig:indexationprocess}. (a) The polar image with the azimuthally integrated pattern plotted on top, as well as the azimuthally integrated best-fit template. (b) The correlation index between azimuthally integrated images and templates plot on the stereographic projection. Features look comparable to the results obtained from the full matching procedure in figure~\ref{fig:indexationprocess} (f).}%
    \label{fig:integratedmatch}
\end{figure}

While the direct approach implementation is reasonably fast, it may still take a few hours to index a dataset of a few gigabytes in size on a common laptop with a limited number of CPU cores.
Therefore, an option was added to pre-filter the template library as described in ref.~\cite{wu2009advances}. 
In this method, the image and templates are integrated over the azimuthal direction, and the correlation between the integrated image and the integrated template library is calculated by a single matrix-vector product.
Only the templates with the highest correlation values are passed onto the full indexation procedure.
In this way the library for full indexation can be significantly reduced, and typically only a tenth of the templates must be preserved to arrive at the same optimum~\cite{wu2009advances}.
However, there is no guarantee that the optimal template will always be included in the pre-filtered set, and trials indicated that the method becomes less reliable for larger template libraries and templates containing more spots.
The integrated matching procedure is illustrated in Figure~\ref{fig:integratedmatch}.

The indexation procedure is then simply repeated for all images in the dataset.
This is parallelized as much as possible using Dask~\cite{rocklin2015dask}.
Dask also makes it possible to process datasets larger than computer memory by only loading parts of the dataset from storage when they need to be processed.

\section{Materials, methods and datasets}
Two datasets were collected inside a JEOL 2200FS TEM operating at 200~kV with a 4K TemCAM-XF416 pixelated CMOS detector (TVIPS).
The first dataset was collected on a sample of Cu-Ag alloy described in ref.~\cite{oellers2020thin}.
The dataset was thoroughly analyzed in ref.~\cite{jeong2021automated}.
The microscope conditions were a beam convergence angle of around 2~mrad, a camera length of 15~cm, a precession angle of 0.3\degree and precession frequency of 100~Hz.
Images were collected at 2k$\times$2k pixel resolution with a dwell time of 50~ms, but were binned to 512$\times$512 pixels for analysis.
The second dataset was collected on a sample of ion irradiated Ti-stabilized stainless steel containing nano-sized precipitates.
The dataset is described in ref.~\cite{cautaerts2021investigation}.
The convergence angle was around 0.5~mrad and the camera length 80~cm; the beam was not precessed.
For both datasets, scan points were about 2~nm apart.
Data was collected in the .tvips file format and converted to the .BLO and .HSPY formats using in house tools that have been made freely available~\cite{niels_cautaerts_2020_4288857}.
Calculations in this paper to generate figures and benchmarks were performed on a desktop computer with a 16 core AMD Ryzen 9 3950x CPU, an NVIDIA RTX 3080 GPU, and 64 GB of RAM running Arch Linux.
The code was also successfully run on a MacBook Pro laptop, a Windows 10 workstation, and the TALOS GPU cluster hosted at the Max Planck Computing \& Data Facility.

\section{Results}

\subsection{Evaluation of reliability}
\label{sec:astarcomp}
To verify the reliability of the implementation, the Cu-Ag dataset was indexed both in Pyxem and using the commercial ASTAR software (NanoMegas).
Indexation results from ASTAR are widely accepted in the community and thus form a suitable benchmark for comparison to the Pyxem method.
In both softwares, the image pre-processing and template simulation parameters were optimized iteratively such that the indexation of randomly chosen diffraction patterns looked reasonable by visual inspection.
In Pyxem, the direct beam in each image was centered, the diffuse background in the images was removed using a difference-of-Gaussians procedure, and the resulting image was smoothed by Gaussian blur.
Thresholding was used to set low intensity pixels to zero and gamma correction with exponent 0.5 was applied to enhance the weaker reflections.
In both ASTAR and Pyxem a template library, derived from the FCC Cu structure, of around 11000 templates was used corresponding to a maximum angular deviation between neighboring orientations of 0.3\degree.
In Pyxem atomic scattering was ignored in simulating the templates and a linear shape function was used for the relrods.
Indexation was performed using the "full indexation" setting in ASTAR and without pre-filtering in Pyxem.
Since ASTAR requires the images to be in 8-bit format, the original 16-bit data was truncated at pixel values of 1000 (the maximum pixel value was around 4000 in the direct beam) and the range 0-1000 was scaled to the 8-bit 0-255 range.
The full Pyxem analysis workflow with parameters for the preceding steps is provided in the supplementary information as a Jupyter notebook.
The original data can be downloaded as a Zenodo dataset.

\begin{figure}[h]
    \centering
    \includegraphics[width=\linewidth]{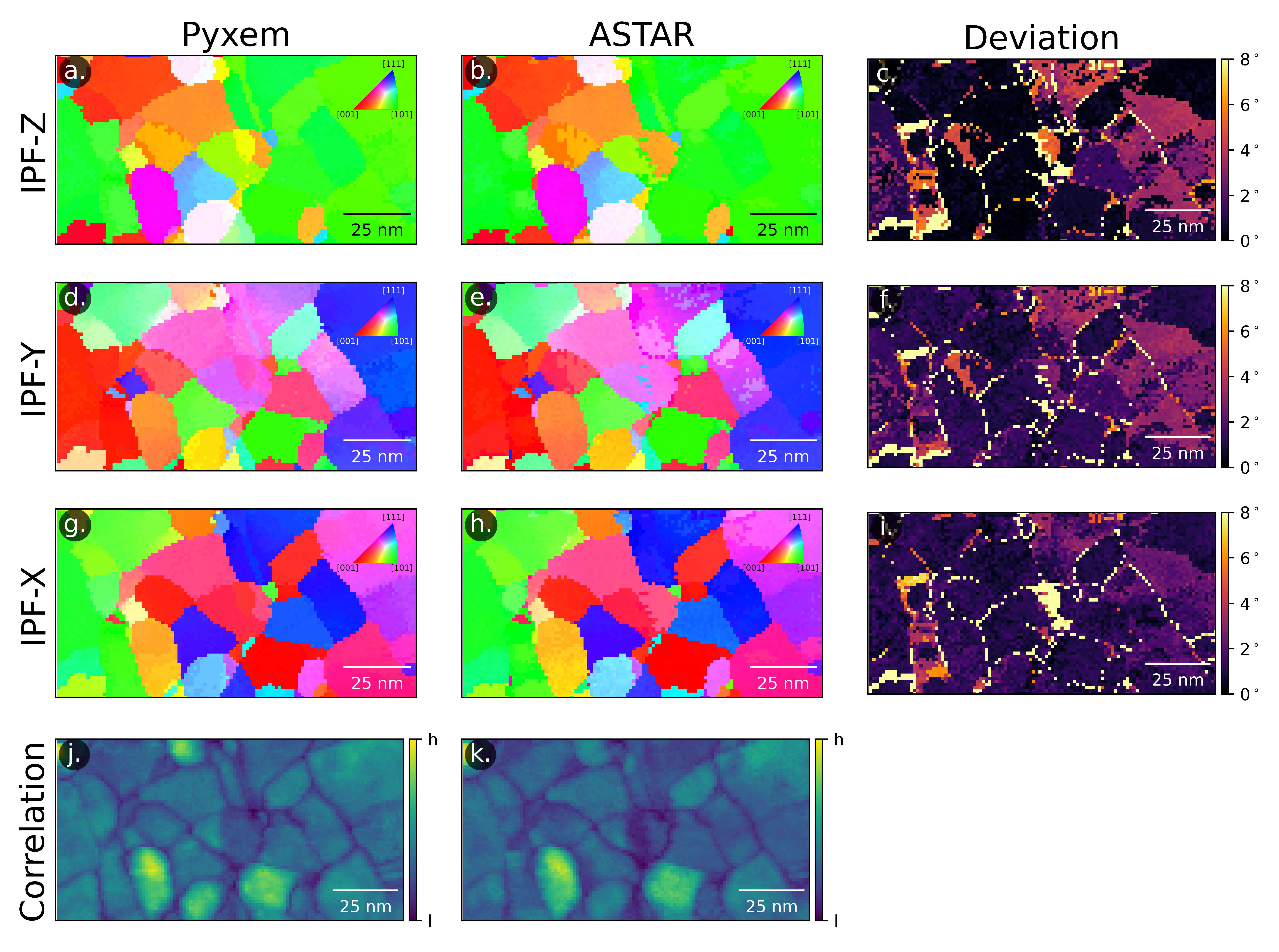}
    \caption{Comparison of the indexation result for Pyxem and ASTAR. (a) and (b) represent the Pyxem and ASTAR IPF-Z maps respectively with (c) the angle between the vectors in each pixel of (a) and (b). Analogous maps are plotted for IPF-Y in (d-f) and IPF-X in (g-i). (j) and (k) show the correlation index maps for Pyxem and ASTAR. Due to differences in intensity range, these maps should not be compared on a pixel-by-pixel basis; only relative intensities within the same map are meaningful.}%
    \label{fig:compareastarglobal}
\end{figure}

The inverse pole figure (IPF) maps for the Z, Y and X direction are shown in Figure~\ref{fig:compareastarglobal} (a), (d) and (g) for Pyxem and Figure~\ref{fig:compareastarglobal} (b), (e), and (h) for ASTAR respectively.
To highlight the differences between the maps, the angular deviation between the ASTAR and Pyxem results are shown in Figure~\ref{fig:compareastarglobal} (c), (f), and (i).
This is calculated as

\begin{equation}
    \theta = \left| \cos^{-1} \left( \frac{v_A \cdot v_P}{|v_A||v_P|} \right) \right|,
\end{equation}

\noindent with $v_A$ and $v_P$ the vector represented in each pixel of the ASTAR and Pyxem maps.

Figure~\ref{fig:compareastarglobal} (j) and (k) show the correlation index maps (equation~\ref{eq:corindex}) for Pyxem and ASTAR respectively.
These two $Q$ maps can not be compared on a pixel by pixel basis because the simulation of templates and normalization are done differently, resulting in different linear scaling.

Correspondence between the two methods is good; for most points in the grain interiors the deviation in grain orientation is less than 3\degree.
The deviation is largest in some pixels near grain boundaries, and in locations where $Q$ is low.
In these regions, there may be ambiguity in the diffraction patterns when multiple crystals contribute to the signal.
Each template in the library can only represent a rotated single crystal, so ambiguous diffraction patterns can be indexed differently depending on small differences between image preprocessing parameters.

\begin{figure}[h]
    \centering
    \includegraphics[width=\linewidth]{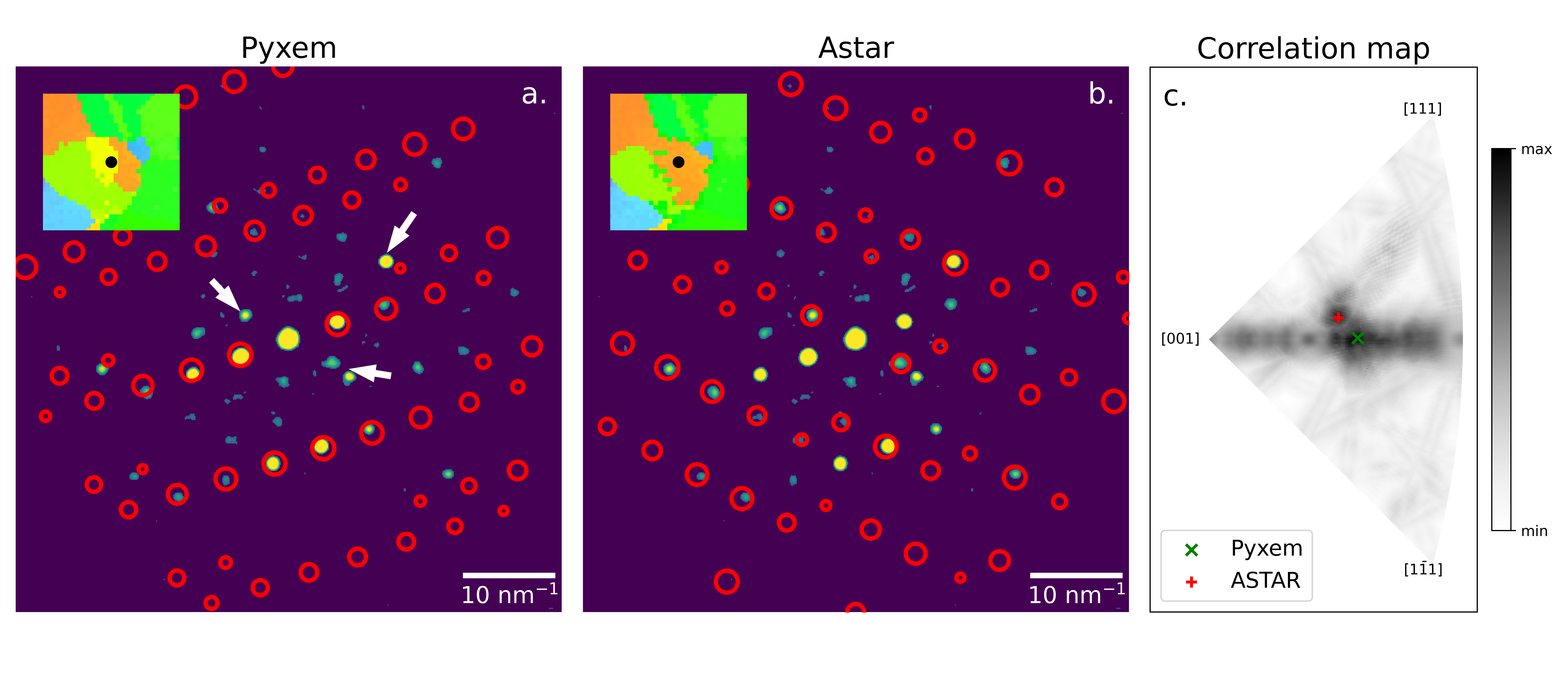}
    \caption{Comparison of the indexation result in (a) Pyxem and (b) ASTAR for a single pattern where there is substantial disagreement. The inset in both images shows the location of the pattern on the IPF-Z map. (c) The correlation map for the pattern obtained with Pyxem, which represents the correlation of each template in the library with the image on the stereographic projection, shows that the Pyxem and ASTAR solutions are local maxima.}%
    \label{fig:compareastarlocal}
\end{figure}

In some locations in the grain interior the two maps differ substantially, as shown for one diffraction pattern in Figure~\ref{fig:compareastarlocal} (a) and (b).
The inset shows the location on the map where the pattern was extracted.
Pixels where the maps differed substantially were associated with ambiguous diffraction patterns containing contributions from overlapping crystals.
The template identified by Pyxem accounts for most of the bright reflections but fails to capture the reflections marked with white arrows that originate from another grain.
The ASTAR indexation does account for these reflections, but ASTAR misses the bright systematic row.
Clearly this is a region where the green grain on the left and the orange grain on the right overlap, resulting in ambiguity.
Indexing overlapping grains is possible in specific cases by using masking techniques~\cite{Valery2017} but they were not implemented for this example.
Figure~\ref{fig:compareastarlocal} (c) shows the correlation map on the stereographic projection as calculated by Pyxem, with the best orientations according to Pyxem and ASTAR indicated.
Both represent local maxima, showing that a minor change in simulation parameters or image processing conditions could change their relative height.

In regions with small deviations between the ASTAR and Pyxem solutions there were no clear metrics to favor one solution over the other, suggesting that for this dataset and method the angular resolution is limited to about 3\degree although this might be improved by additional image processing~\cite{jeong2021automated}.

\subsection{Performance}

\begin{figure}[h]
    \centering
    \includegraphics[width=\linewidth]{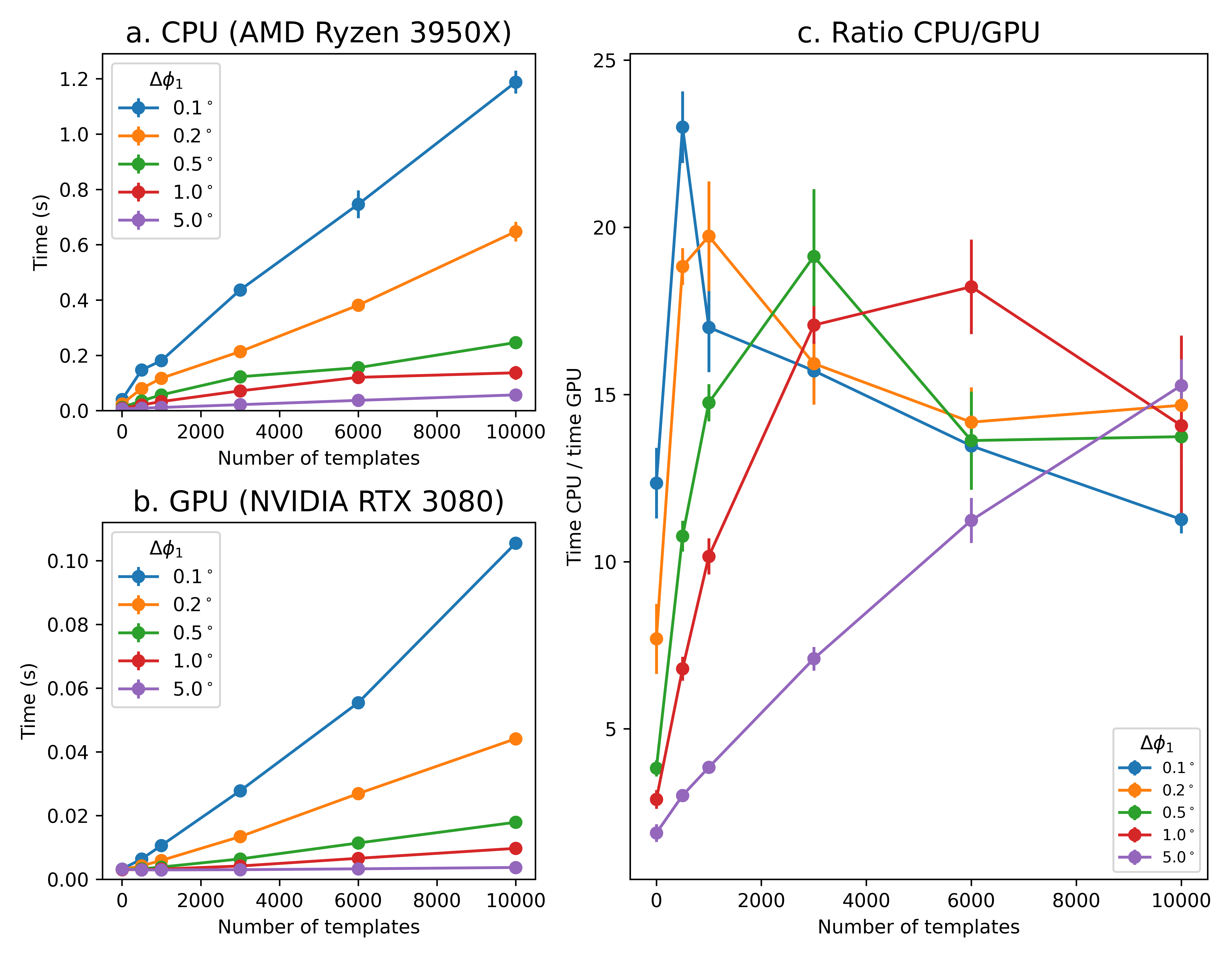}
    \caption{Time to index a single diffraction pattern image of size $256 \times 256$ for various angular increments and library sizes, (a) timed on a 16 core CPU and (b) timed on a RTX 3080 GPU. Note the differences in the y-axis scales. (c) The ratio of CPU time over GPU time shows the substantial advantage of the GPU.}%
    \label{fig:benchmarkoneim}
\end{figure}

The time to index a single pattern depends on the $\Delta \phi_1$ sampling, the number of templates in the library, the number of spots in the templates, and the available computational resources.
Figure~\ref{fig:benchmarkoneim} shows the time it took to index a single 256$\times$256 pattern for different library sizes and $\Delta\phi_1$ on (a) a 16 core CPU and (b) an NVIDIA RTX 3080 GPU in a desktop computer.
Each point represents the mean of 10 independent measurements, the error bar represents $\pm$ the standard deviation.
The CPU timing includes the time to convert the image to polar coordinates, to filter the original template library of about 11000 templates down to the desired number of templates, and to find the best in-plane rotation for the remaining templates.
The GPU timing also includes the time to transfer the image to the GPU and to send the result back.

To a good approximation, the computational time increases linearly with the number of templates in the library and with $1/\Delta \phi_1$.
On both the CPU and GPU, the y-intercept of the lines is close to zero, showing that the template matching step, including the in-plane angle optimization, dominates the computational time for most conditions.
Figure~\ref{fig:benchmarkoneim} (c) shows the ratio of the CPU time and the GPU time.
On this system, the GPU was about 10-15 times faster than the CPU for most conditions.
Experiments on a dual core Macbook laptop and a TALOS cluster node with 40 cores showed that speed scales roughly linearly with number of cores; therefore, a lot of CPU cores are necessary before the performance of a mid-range GPU is matched. 
The relative benefit of the GPU decreases with smaller template libraries and larger increments of $\Delta \phi_1$, since the relative time cost of data transfers to and from the GPU increases.
The acceleration of the polar transformation and library filter steps still make the GPU outperform the CPU.

The time necessary to index an entire dataset can be roughly extrapolated from Figure~\ref{fig:benchmarkoneim} by multiplying by the number of patterns.
For example, indexing the entire dataset from Figure~\ref{fig:compareastarglobal} with 11000 templates and without template pre-filtering took around 3 minutes on the GPU (20~ms per pattern) and about 16 minutes on the CPU (110~ms per pattern).
The time the process takes on the GPU is a factor 2 slower than expected, which can be attributed to the fact that the image pre-processing is also included in the timing and these steps are performed on the CPU.
Some benchmarking results from the Macbook and from the Talos cluster are given in the supplementary materials.

\subsection{Custom indexation procedures}
\begin{figure}[h!]
    \centering
    \includegraphics[width=0.9\linewidth]{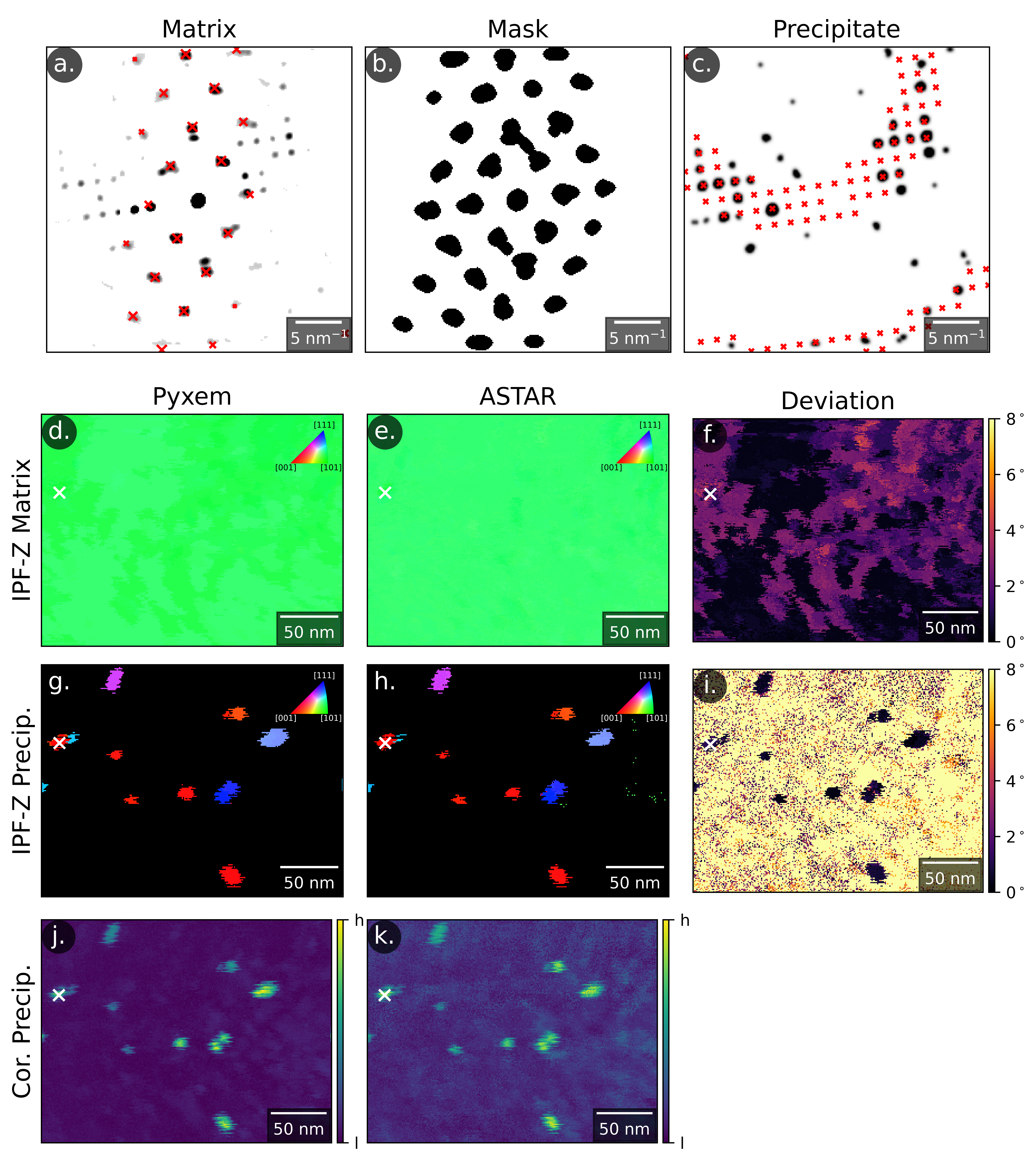}
    \caption{Illustration of the two-stage indexation procedure. The dataset is first indexed with the matrix phase as shown in (a). (b) A mask is created to index the templates by averaging all the images and thresholding. (c) The masked dataset is indexed using the template library from the precipitates. (d), (e), and (f) show the IPF-Z maps obtained from indexing the matrix phase with Pyxem, ASTAR, and the angular deviation between them, respectively. (g) (h) and (i) show analogous maps for the precipitates (low correlation index pixels were set to black). (j) and (k) are the Pyxem and ASTAR correlation index maps for the precipitate indexation.}%
    \label{fig:gphase}
\end{figure}

Thanks to the flexibility of Pyxem, customized workflows can easily be constructed.
This section illustrates a two-stage indexing procedure for identifying orientations of overlapping phases, specifically for the case of nano-sized precipitates embedded in a crystalline matrix.
Since the reflections of the matrix phase always dominate over the signal of the precipitates, a reliable indexation of the precipitates requires a subtraction of the matrix signal from the images.
Strategies to achieve this are described in refs.~\cite{Valery2017, Rauch2019} and the tools were successfully applied to reveal orientations of nanoprecipitates in irradiated steel~\cite{Cautaerts2020}.
The technique consists of three stages: first the matrix is indexed, then the dominant matrix contribution in the experimental patterns is masked, and finally the precipitate fraction is indexed on the masked dataset.
This is illustrated for one pattern in the irradiated steel dataset from ref.~\cite{Cautaerts2020} in Figure~\ref{fig:gphase} (a), (b) and (c).
The tools that were used to analyze this data in ref.~\cite{Cautaerts2020} are proprietary, and as of yet not widely accessible.

Here, a similar, fully reproducible workflow was constructed using Pyxem to analyze the same dataset.
The complete data processing pipeline is provided in a Jupyter notebook in the supplementary materials.
As with the Cu-Ag dataset, the patterns were centered and background-subtracted.
Subsequently an affine transformation was applied to the images to correct for projector astigmatism.
The elements of the affine transformation matrix were iteratively optimized based on best fit templates of individual patterns.
The patterns were subsequently indexed using a library of about 1000 templates (maximum deviation of 1\degree between templates) from FCC Fe.
The result of the indexation is shown in Figure~\ref{fig:gphase} (d) as an IPF-Z map, showing that the data was collected in a single grain close to a $\langle 110 \rangle$ zone.
Figure~\ref{fig:gphase} (e) and (f) show the result from ref.~\cite{Cautaerts2020} and its angular deviation from the Pyxem result respectively.
Deviations up to about 4\degree can be observed.

To remove the intense matrix reflections, a mask was created by calculating the average of all the diffraction patterns and thresholding the resulting image.
A number of erosion and dilation operations were performed to remove small noise-like features in the mask, and to ensure the removal of tails from intense diffraction peaks.
The final mask that was applied to the images is shown in Figure~\ref{fig:gphase} (b).

The masked dataset was subsequently indexed using a library of about 4000 templates (maximum deviation of 0.5\degree between templates) from Mn$_6$Ni$_{16}$Si$_7$ phase (G-phase).
Due to the larger lattice parameter of the G-phase compared to the matrix phase (austenite), templates contain many more reflections and small changes in orientation can drastically change the patterns.
For this reason, a larger template library was simulated using smaller orientation increments.
Additionally, a very small relrod width was chosen, such that only the reflections very close to the Ewald sphere would be included.

The parameters of the indexation procedure were optimized iteratively on individual patterns containing signal from the precipitates.
This process showed the necessity of introducing two somewhat counter-intuitive processing steps: firstly, the intensity of all diffraction spots in the template library were set to unity (ignoring both the structure factor and relrod shape factor), and secondly a small constant (about 10\% of the maximum pixel intensity) was subtracted from all pixels in the experimental images.
Without these steps many patterns were erroneously indexed, because intense reflections in the template dominated in the evaluation of $Q$.
After the gamma correction and the masking, the relative intensities of the weak spots in the experimental images were no longer relevant to determining the orientation.
Instead, the best indexation represents the template with the highest number reflections matching spots in the image and the fewest reflections not matching any spots.
By setting all intensities in the template to unity, each reflection has equal weight in the calculation of $Q$, and by subtracting a small value from each pixel in the image, the reflections of the template that do not match a spot provide a negative contribution to $Q$.

To index the G-phase these steps are acceptable, since higher order Laue zones and the curvature of the Ewald sphere are visible in most diffraction patterns (see Figure~\ref{fig:gphase} (b)). 
These features are very sensitive to small changes in orientation.
Figure~\ref{fig:gphase} (g) shows the indexed result from the masked dataset as an IPF-Z map; pixels with low correlation index were set to black to reveal the precipitates.
The correlation index map is shown in Figure~\ref{fig:gphase} (j).
The orientation and correlation maps from ref.~\cite{Cautaerts2020} are shown in Figure~\ref{fig:gphase} (h) and Figure~\ref{fig:gphase} (k) respectively.
The angular deviation between the Pyxem and proprietary solutions is given in Figure~\ref{fig:gphase} (i).
Within the particles, deviation between the solutions is less than 1\degree, outside the particles the indexation is arbitrary since there is only noise in these patterns. 
Because the indexation only relies on the presence of spots and not their intensity, and because all templates in the library are very different from each other due to the small relrod width, the orientation of the precipitates is determined with very high precision.
In the template library of the matrix phase, many templates are very similar and only differ in spot intensity, which results in a lower precision in the indexation.
A smaller camera length, such that spots from higher order Laue zones would also be captured in the image, may improve the indexing precision of the matrix.

Alternative approaches to indexing the dataset were also attempted.
For example, instead of using a single mask, the best matching template from each pattern in the first indexation step can be converted to a mask by placing circles at the coordinates of each reflection.
For this particular dataset this proved to be an inferior strategy because there were features in the diffraction pattern that did not correspond to either of the phases, instead originating from defects or surface contamination and oxidation.
These features were not masked with the template mask approach and subsequently interfered with the indexation of the precipitates.
However, this type of approach may be necessary for datasets of polycrystalline materials containing multiple phases, since the required mask is different in every grain.
Finally, it may also be possible to separate contributions from different phases in the diffraction patterns using unsupervised learning techniques, such as principal component analysis (PCA) and non-negative matrix factorization (NMF)~\cite{martineau2019unsupervised}.
These techniques are also implemented in Pyxem and may serve as a helpful preprocessing step before template matching.

\section{Discussion}
This paper demonstrated the template matching capabilities in Pyxem for extracting orientation information from NBED and PED datasets.
Two datasets were provided as examples, illustrating the speed and extensibility of the solution.
Accuracy was benchmarked against commercial solutions; deviations between the solutions could be attributed to ambiguity in the diffraction patterns from overlapping crystals or limited differentiability among templates in the library.
Under these circumstances, small differences in image pre-processing or library simulation conditions could have an outsized effect on the best matching template for ambiguous patterns.
With the integrated workflow offered by Pyxem, it is relatively easy to investigate the effect of these pre-processing parameters and their relative importance to the result.
From this analysis, the maximum precision of the indexation result can be estimated.
While 1\degree~is the often quoted maximum precision of NBED/PED orientation mapping~\cite{Zaefferer2011, morawiec2014orientation}, as was shown in this paper the precision can vary strongly depending on experimental parameters such as camera length, local orientation and lattice parameters of the phases involved.
In general, if the differentiation between neighboring templates relies primarily on differences in spot intensity the uncertainty could be as high as $\pm$5\degree; the presence of reflections from higher order Laue zones strongly improves indexing reliability.

Other advantages offered by Pyxem are that all file formats supported by HyperSpy can be analyzed.
Data can be preprocessed using any of the tools in the scientific python ecosystem and integrated into a fully reproducible analysis pipeline. 
Template libraries can be generated in the same environment and freely modified using the diffsims library, and large datasets can be indexed with large template libraries in a short amount of time thanks to GPU support.
Workflow flexibility, openness, speed, and reproducibility are becoming increasingly important in light of the growing popularity of fast pixelated detectors which are producing ever more complex datasets.

The open source nature of the code means that it can be freely adapted and extended, for example towards real-time indexing.
A single diffraction pattern can be reliably indexed within a few to tens of milliseconds, which is on the same order as the acquisition time for a pattern on a CMOS or CCD based detector.
Hence these processes could easily run in parallel meaning that routine orientation mapping of single phase materials with NBED could become automated like electron backscatter diffraction (EBSD) in the scanning electron microscope.
A prototype along these lines is being worked out by integrating with the LiberTEM library~\cite{clausen2020libertem, clausen-alexander-2020-3982290}.
For the extremely fast direct electron detectors, such as the EMPAD detector, the code would not be fast enough to perform real time indexing.
Multiple optimizations could still be explored, like improving the algorithm for optimizing the in-plane angle $\phi_1$ and parallelizing the image preprocessing steps.

\section{Conclusion}
In this paper, the new template matching capabilities implemented in Pyxem for analyzing NBED and precession diffraction data were discussed, from implementation, through performance, to applications.
The code enabled reliable and fast orientation mapping in nanostructured materials on high quality datasets originating from a pixilated detector.
Through the fine grained control over the entire analysis process the effects of image processing and template simulation parameters on the result could be evaluated.
In addition, a two stage indexation procedure was implemented for mapping the orientation of nano-sized precipitates embedded in a crystalline matrix, a workflow that could be adapted by many other researchers.
All the workflows in this paper are fully documented and reproducible and can serve as templates for other open, custom and transparent analysis workflows.

\section*{Acknowledgements}
We acknowledge funding through BiGmax (https://www.bigmax.mpg.de/), the Max Planck research network on big-data-driven materials science.
We also want to sincerely thank all the contributors to HyperSpy and Pyxem, who have dedicated their time to building free analysis libraries that will benefit all future researchers and lay the foundation for open science in electron microscopy.
In particular we wish to acknowledge the original creator of the Pyxem library D. Johnstone, who's pioneering work this work built on.
P.C. acknowledges the support of his studentship from the EPSRC.
H.W.{\AA} acknowedges NTNU for financial support through the NTNU Aluminium Product Innovation Centre.

\bibliographystyle{model1a-num-names}
\bibliography{references}


\end{document}